\documentclass{elsart}
\usepackage{thophys}
\usepackage{graphics,feynmp}
\usepackage{amsmath,amssymb}

\DeclareMathOperator{\astcomma}{\stackrel{\ast}{,}}
\newcommand{\ii}{\mathrm{i}}
\newcommand{\ee}{\mathrm{e}}
\newcommand{\dd}{\mathrm{d}}
\allowdisplaybreaks
\makeindex
\begin{document}
\begin{frontmatter}
\title{%
  Unitarity of Time-Like Noncommutative~Gauge~Theories:\\
  The Violation of Ward~Identities in
  Time-Ordered~Perturbation~Theory\thanksref{PPN}}
\runauthor{Ohl, R\"uckl and Zeiner}
\author{Thorsten Ohl}
\ead{ohl@physik.uni-wuerzburg.de}
\author{Reinhold R\"uckl}
\ead{rueckl@physik.uni-wuerzburg.de}
\author{J\"org Zeiner}
\ead{zeiner@physik.uni-wuerzburg.de}
\address{%
  Institut~f\"ur~Theoretische~Physik~und~Astrophysik,
  Universit\"at~W\"urzburg,
  Am~Hubland,
  97074~W\"urzburg,
  Germany}
\begin{abstract}
  We study Ward identities for simple processes with external gauge
  bosons in the time-ordered perturbation theory approach to time-like
  noncommutative gauge theories.  We demonstrate that these Ward
  identities cannot be satisfied
  when all orders in the noncommutativity parameters~$\theta_{i0}$ are taken
  into account.  We conclude that in time-ordered perturbation theory
  one cannot solve the unitarity problem of time-like
  noncommutative quantum field theories.
\end{abstract}
\thanks[PPN]{WUE-ITP-2003-005 [hep-th/0309021] (September 2003)}
\end{frontmatter}
\begin{fmffile}{\jobname pics}
\fmfset{arrow_ang}{10}
\fmfset{curly_len}{2mm}
\fmfset{wiggly_len}{3mm}
\section{Introduction}
Noncommutative Quantum Field Theory~(NCQFT) has recently received
renewed attention (see~\cite{Douglas:2001ba} for a review).  This
interest is triggered by the connection of NCQFT to string theory, by
its nature as a non-local generalization of Quantum Field
Theory~(QFT) and by the possibility of experimental tests, provided
the scale of the noncommutativity is sufficiently small.

NCQFT starts from the assumption that the familiar continuous
Minkowski space-time with coordinates~$x_\mu$ is the long-distance
limit of a space-time geometry with noncommuting coordinates~$\hat x_\mu$,
satisfying commutation relations
\begin{equation}
\label{eq:NC}
   [\hat x_\mu, \hat x_\nu] = \frac{\ii\theta_{\mu\nu}}{\Lambda^2}\,.
\end{equation}
In general, the antisymmetric matrix~$\theta_{\mu\nu}$ may depend
on~$\hat x_\mu$.  In this note, like in most previous studies,
$\theta_{\mu\nu}$ is assumed to be constant.  The noncommuting
coordinates can be realized on an ordinary commuting space-time by the
associative Moyal $\ast$-product
\begin{equation}
\label{eq:Moyal}
  (f\ast g) (x) = \lim_{\xi,\eta\to0} \left[
     \ee^{\ii\partial_{\xi}\wedge\partial_{\eta}}
     f(x+\xi) g(x+\eta)\right]
\end{equation}
which in momentum space induces phase factors~$\ee^{\ii p\wedge q}$ 
with the antisymmetric product
\begin{equation}
\label{eq:wedge}
  p\wedge q = \frac{1}{2\Lambda^2} p^\mu \theta_{\mu\nu} q^\nu\,.
\end{equation}
In order to construct a NCQFT from a given QFT one can make use of
a correspondence principle according to which all ordinary products of
fields in the Lagrangian are replaced by Moyal
$\ast$-products (see~\cite{Douglas:2001ba} and references cited
therein).  As a result, all interaction vertices acquire
momentum-dependent phase factors.

The Moyal $\ast$-product~(\ref{eq:Moyal}) involves derivatives of all
orders and thus makes the theory non-local.  Great care has therefore to be
taken that the physical interpretation of the theory is not spoiled by
this non-locality.  While it appears that the non-locality can be
controlled for space-like noncommutativity,
i.\,e.~$\theta_{ij}\not=0$~\cite{Alvarez-Gaume:2003mb}, there are 
serious problems with unitarity in the case of time-like
noncommutativity, i.\,e.~$\theta_{i0}\not=0$, when all orders
in~$\theta_{i0}$ are taken into account.  The perturbative
$S$-matrix is no longer unitary, since the cutting rules are violated,
as shown by an explicit calculation in the usual covariant
perturbation theory~\cite{Gomis:2000zz}.

In~\cite{Liao:2002xc}, Time-Ordered Perturbation Theory~(TOPT) for
NCQFT was introduced as an attempt to solve the unitarity
problem~\cite{Gomis:2000zz} by constructing a manifestly unitary time
evolution operator in non-covariant perturbation theory. 
In~\cite{Liao:2002pj}, it has then been demonstrated that TOPT
indeed solves the unitarity
problem for scalar fields in time-like NCQFT. 
Applications of TOPT to processes without external gauge bosons in
Noncommutative Quantum Electrodynamics~(NCQED) have been discussed
in~\cite{Liao:2002kd}.  In section~\ref{sec:TOPT} of the present
article, we briefly review TOPT in order to establish
our notation and complete the Feynman rules given
in~\cite{Liao:2002xc,Liao:2002kd}.

However, the formal unitarity of the time evolution operator
does not suffice in covariantly quantized
gauge theories, where unphysical degrees of
freedom contribute negative norm states.  The unphysical degrees of
freedom must be proven to decouple from the $S$-matrix to obtain
unitarity in the physical subspace.  In sections~\ref{sec:WI}
and~\ref{sec:no-go} we will
show that the tree-level Ward Identities~(WIs)
\begin{equation}
\label{eq:WI}
  \frac{\partial}{\partial x^\mu}
    \Braket{0|\mathrm{T} A^\mu(x)
       \Phi_1(x_1)\Phi_2(x_2)\cdots\Phi_n(x_n)|0}_{\text{amputated, on-shell}} = 0
\end{equation}
with external photons cannot be satisfied
in TOPT for time-like NCQED.  The Green function~(\ref{eq:WI}) is
amputated for all fields~$\Phi_i$.
The amputated gauge field~$A_\mu$ corresponds to a current insertion.

A practical consequence of the violation of the WIs~(\ref{eq:WI}) is
that differential cross sections calculated with covariant
polarization sums---including sums over ghosts---are no longer
positive definite.  Indeed, we were alerted to the problem under investigation
by this phenomenon in the calculation of Compton scattering.  Furthermore, the
violation of tree level WIs with external gauge bosons will obviously
invalidate the cutting rules for loops involving gauge bosons.

Thus TOPT does not solve the unitarity problem in NCQED for
processes with external photons.    Our proof of this statement carries
over to any time-like Noncommutative Gauge Theory~(NCGT).

\section{Time-Ordered Perturbation Theory}
\label{sec:TOPT}
Upon integration over space-time, the $\ast$-product~(\ref{eq:Moyal})
is cyclically symmetric and reduces to the ordinary product for one
pair of factors
\begin{subequations}
\label{eq:space-time-integral}
\begin{align}
\label{eq:Moyal-cyclicity}
  \int\!\dd^4x\,(f_1\ast f_2\ast\cdots\ast f_n)(x)
      &= \int\!\dd^4x\,(f_2\ast\cdots\ast f_n\ast f_1)(x)\\
\label{eq:Moyal-binary}
  \int\!\dd^4x\,(f_1\ast\cdots\ast f_n)(x)
      &= \int\!\dd^4x\,(f_1\ast\cdots\ast f_i)(x)(f_{i+1}\ast\cdots\ast f_n)(x) \,.
\end{align}
\end{subequations}
This can be seen by partial integration, using the antisymmetry
of (\ref{eq:wedge}).
It follows from~(\ref{eq:Moyal-binary}) that the propagators remain
unchanged when ordinary products are replaced by $\ast$-products.
In addition, (\ref{eq:Moyal-cyclicity}) shows that the additional
phases are invariant under cyclic permutations.

In~\cite{Liao:2002xc} it has been proposed to use TOPT to construct formally
unitary quantum field theories with time-like noncommutativity.  In
TOPT, each propagator is split into a positive energy and
negative energy piece:
\begin{equation}
\label{eq:TOPT}
  \parbox{34mm}{%
    \fmfframe(7,5)(7,5){%
      \begin{fmfgraph*}(20,25)
        \fmfleft{p1,p2,dl}
        \fmfright{dr,p1',p2'}
        \fmflabel{$p_1$}{p1}
        \fmflabel{$p_2$}{p2}
        \fmflabel{$p_1'$}{p1'}
        \fmflabel{$p_2'$}{p2'}
        \fmf{plain}{p1,v,p2}
        \fmf{plain}{p1',v',p2'}
        \fmf{plain}{v,v'}
        \fmfdot{v,v'}
      \end{fmfgraph*}}} =
  \parbox{34mm}{%
    \fmfframe(7,5)(7,5){%
      \begin{fmfgraph*}(20,25)
        \fmfleft{p1,p2,dl}
        \fmfright{dr,p1',p2'}
        \fmflabel{$p_1$}{p1}
        \fmflabel{$p_2$}{p2}
        \fmflabel{$p_1'$}{p1'}
        \fmflabel{$p_2'$}{p2'}
        \fmf{plain}{p1,v,p2}
        \fmf{plain}{p1',v',p2'}
        \fmf{dashes,label=$q^{(+)}$,label.dist=3pt}{v,v'}
        \fmffixed{(.25w,.8h)}{v,v'}
        \fmfdot{v,v'}
        \fmfforce{(whatever,-.2h)}{t1}
        \fmfforce{(whatever,1.0h)}{t2}
        \fmffixedx{0}{t1,v}
        \fmffixedx{0}{t2,v}
        \fmfforce{(whatever,-.2h)}{t1'}
        \fmfforce{(whatever,1.0h)}{t2'}
        \fmffixedx{0}{t1',v'}
        \fmffixedx{0}{t2',v'}
        \fmf{dots}{t1,t2}
        \fmf{dots}{t1',t2'}
        \fmflabel{$t'$}{t2'}
        \fmflabel{$t$}{t2}
      \end{fmfgraph*}}} +
  \parbox{34mm}{%
    \fmfframe(7,5)(7,5){%
      \begin{fmfgraph*}(20,25)
        \fmfleft{p1,p2,dl}
        \fmfright{dr,p1',p2'}
        \fmflabel{$p_1$}{p1}
        \fmflabel{$p_2$}{p2}
        \fmflabel{$p_1'$}{p1'}
        \fmflabel{$p_2'$}{p2'}
        \fmf{plain}{p1,v,p2}
        \fmf{plain}{p1',v',p2'}
        \fmf{dashes,label=$q^{(-)}$,label.side=right,label.dist=3pt}{v,v'}
        \fmffixed{(-.25w,.8h)}{v,v'}
        \fmfdot{v,v'}
        \fmfforce{(whatever,-.2h)}{t1}
        \fmfforce{(whatever,1.0h)}{t2}
        \fmffixedx{0}{t1,v}
        \fmffixedx{0}{t2,v}
        \fmfforce{(whatever,-.2h)}{t1'}
        \fmfforce{(whatever,1.0h)}{t2'}
        \fmffixedx{0}{t1',v'}
        \fmffixedx{0}{t2',v'}
        \fmf{dots}{t1,t2}
        \fmf{dots}{t1',t2'}
        \fmflabel{$t$}{t2'}
        \fmflabel{$t'$}{t2}
      \end{fmfgraph*}}}
\end{equation}
Here all four-momenta are taken on-shell, in particular
\begin{equation}
\label{eq:on-shell}
  q^{(\pm)}=(\pm\sqrt{{\vec q}^2+m^2},\vec q)\,.
\end{equation}
Since the three-momentaare conserved at a vertex, 
$\vec q=\vec p_1+\vec p_2=\vec p_1'+\vec p_2'$,
energy cannot be conserved.

After Fourier transformation~($\mathrm{F.T.}$), the Moyal
phase~$\varphi$ of a $n$-point vertex
\begin{equation}
\label{eq:vertex}
  (\Phi_1\ast\Phi_2\ast\cdots\ast\Phi_n)(x)
    \stackrel{\mathrm{F.T.}}{\longrightarrow}
      \ee^{-\ii\varphi(p_1,p_2,\ldots,p_n)}
         \Phi_1(p_1)\Phi_2(p_2)\cdots\Phi_n(p_n)
\end{equation}
is given by
\begin{equation}
\label{eq:phi}
  \varphi(p_1,p_2,\ldots,p_n) = \sum_{i<j} p_i\wedge p_j
\end{equation}
using the notation~(\ref{eq:wedge}).  The phase~$\varphi$ defined
in~(\ref{eq:phi}) is not cyclically symmetric
\begin{equation}
\label{eq:non-cyclic}
  \varphi(p_1,p_2,\ldots,p_n)\not=\varphi(p_2,\ldots,p_n,p_1)\,.
\end{equation}
Only in the case of four-momentum conservation,
the contributions from either the first or the last momentum cancel
\begin{equation}
\label{eq:Moyal-reduced}
  \varphi(p_1,p_2,\ldots,p_n)\Big|_{p_1+p_2+\cdots+p_n=0}
    = \varphi(p_1,p_2,\ldots,p_{n-1})
    = \varphi(p_2,\ldots,p_n)
\end{equation}
and cyclical symmetry is recovered.  Therefore,
(\ref{eq:non-cyclic})~introduces an ordering ambiguity
in~(\ref{eq:vertex}) that must be taken into account.

Using the notation~$q^{(\pm)}_0=\pm\sqrt{{\vec q}^2 + m^2}$,
the two scalar propagators in TOPT can be expressed as~\cite{Liao:2002xc}
\begin{equation}
\label{eq:scalar-TOPT}
  \parbox{40mm}{%
    \fmfframe(4,5)(4,5){%
      \begin{fmfgraph*}(30,5)
        \fmfleft{l}
        \fmfright{r}
        \fmf{dashes,label=$q^{(\lambda)}$}{l,r}
      \end{fmfgraph*}}}
     = \frac{\ii}{2q^{(\lambda)}_0}
    \frac{1}{q_0-q^{(\lambda)}_0+\lambda\ii\epsilon}\,.
\end{equation}
Since the phase~(\ref{eq:phi}) depends
on all four-momenta at the vertex, it is generally
impossible to remove the dependence of~$\varphi$ on the two on-shell
momenta~$q^{(\pm)}$ with positive and negative energy.  Consequently, the two
contributions to the scattering process in~(\ref{eq:TOPT})
have different Moyal phases~$\phi(q^{(\pm)})$, where the
dependence on the external four-momenta is suppressed for brevity.
After adding up these two contributions, the same poles
as in covariant perturbation theory appear
\begin{equation}
\label{eq:poles}
  \sum_{\lambda=\pm}
      \ee^{\ii\phi(q^{(\lambda)})}
      \frac{1}{2q^{(\lambda)}_0} \frac{1}{q_0-q^{(\lambda)}_0+\lambda\ii\epsilon}
    = \frac{R(q^{(+)},q^{(-)})}{q_0^2 - \left.\vec q\right.^2 - m^2 + \ii\varepsilon}
    = \frac{R(q^{(+)},q^{(-)})}{q^2 - m^2 + \ii\varepsilon}\,,
\end{equation}
however the residue is not simply unity, but a linear combination of
the phase factors~$\ee^{\ii\phi(q^{(\pm)})}$~\cite{Liao:2002xc}:
\begin{equation}
\label{eq:R}
  R(q^{(+)},q^{(-)}) =
    \frac{1}{2}\sum_{\lambda=\pm}
       \ee^{\ii\phi(q^{(\lambda)})}\left(1+\frac{q_0}{q^{(\lambda)}_0}\right)\,.
\end{equation}
At this point we want to stress that the structure of~(\ref{eq:poles})
allows the following approach to WIs: in a first step, we can ignore
the phases from TOPT and draw conclusions from the required cancellation
of poles alone.  In the second step, we can then use the results from
the first requirement and derive relations among the phases.

The authors of~\cite{Liao:2002xc} have not given a prescription for
propagators with momenta in the numerator.  This will be done here.
The propagator for a spin-$1/2$ field, is composed of the positive and
negative energy contributions
\begin{subequations}
\begin{align}
   S^{(+)}(x)
     &= \int\frac{\dd^3\vec q}{(2\pi)^3 2q^{(+)}_0}
          \sum_{s=\pm} u_s(q)\bar u_s(q) e^{-\ii q^{(+)}x} \notag \\
     &= \int\frac{\dd^3\vec q}{(2\pi)^3 2q^{(+)}_0}
          \left(\fmslash{q}^{(+)} + m\right) e^{-\ii q^{(+)}x} \\
\label{eq:S(-)}
   S^{(-)}(x)
     &= \int\frac{\dd^3\vec q}{(2\pi)^3 2q^{(+)}_0}
          \sum_{s=\pm} v_s(q)\bar v_s(q) e^{\ii q^{(+)}x}
      = \int\frac{\dd^3\vec q}{(2\pi)^3 2q^{(+)}_0}
          \left(\fmslash{q}^{(+)} - m\right) e^{\ii q^{(+)}x} \notag \\
     &= \int\frac{\dd^3\vec q}{(2\pi)^3 2q^{(-)}_0}
         \left(\fmslash{q}^{(-)} + m\right) e^{-\ii q^{(-)}x}\,,
\end{align}
\end{subequations}
where we have substituted~$\vec q\to-\vec q$ in the last step
of~(\ref{eq:S(-)}).  Hence the spin-$1/2$ propagators
in momentum space in TOPT are given by
\begin{equation}
\label{eq:fermion-TOPT}
  \parbox{40mm}{%
    \fmfframe(4,5)(4,5){%
      \begin{fmfgraph*}(30,5)
        \fmfleft{l}
        \fmfright{r}
        \fmf{fermion,label=$q^{(\lambda)}$}{l,r}
      \end{fmfgraph*}}}
    = \frac{\ii}{2q^{(\lambda)}_0}
      \frac{\fmslash{q}^{(\lambda)}+m}{q_0-q^{(\lambda)}_0+\lambda\ii\epsilon}\,.
\end{equation}
Just as for scalar particles, the poles are the same in TOPT as in covariant
perturbation theory, but the residue is modified and a regular
term is added:
\begin{multline}
\label{eq:poles-spin-1/2}
  \sum_{\lambda=\pm}
    \ee^{\ii\phi(q^{(\lambda)})}
    \frac{1}{2q^{(\lambda)}_0}
    \frac{\fmslash{q}^{(\lambda)}+m}{q_0-q^{(\lambda)}_0+\lambda\ii\epsilon} =
  \frac{1}{q^2-m^2+\ii\varepsilon} \times \\ \Biggl(
          R_+(q^{(+)},q^{(-)})\left(\fmslash{q}+m\right)
        + R_-(q^{(+)},q^{(-)})
             \left(   \gamma^0 q^{(+)}_0
                    - \frac{q_0}{q^{(+)}_0}(\vec\gamma\vec q - m) \right) \Biggr) \\
  = \frac{R(q^{(+)},q^{(-)})}{\fmslash{q}-m+\ii\varepsilon}
      - \gamma^0\frac{R_-(q^{(+)},q^{(-)})}{q^{(+)}_0}
\end{multline}
with~$R$ from~(\ref{eq:R}) and
\begin{equation}
\label{eq:R+/-}
  R_{\pm}(q^{(+)},q^{(-)}) = \frac{1}{2}
    \left( \ee^{\ii\phi(q^{(+)})} \pm \ee^{\ii\phi(q^{(-)})} \right)\,.
\end{equation}
The gauge boson propagator in TOTP can be derived
analogously.  For our purpose it suffices to consider only
Feynman gauge ($\xi=1$), in which case there are no momenta in the
numerator and the propagator is given by
\begin{equation}
\label{eq:photon-TOPT}
  \parbox{40mm}{%
    \fmfframe(4,5)(4,5){%
      \begin{fmfgraph*}(30,5)
        \fmfleft{l}
        \fmfright{r}
        \fmf{boson,label=$q^{(\lambda)}$}{l,r}
      \end{fmfgraph*}}}
    = \frac{-\ii g_{\mu\nu}}{2q^{(\lambda)}_0}
      \frac{1}{q_0-q^{(\lambda)}_0+\lambda\ii\epsilon}\,.
\end{equation}

\subsection{NCQED}
\label{sec:NCQED}
The arguments that will be put forward in section~\ref{sec:WI} are
valid for arbitrary~$\mathrm{U}(N)$ NCGTs.  For $N=1$ all nonzero
charges have to be the same, up to a sign.
NCGTs for~$\mathrm{SU}(N)$ and $\mathrm{U}(1)$ with different
nonzero charges can be constructed in the enveloping
algebra~\cite{Wess:pr}.
The investigation of the interplay of gauge invariance, temporal
non-locality and unitarity in the formulation~\cite{Wess:pr} requires
the study of the corresponding Seiberg-Witten maps to all orders
in~$\theta_{\mu\nu}$ and will be the subject of future research.
For simplicity, we confine ourselves to a $\mathrm{U}(1)$-theory with
one massive spin-$1/2$ matter field (NCQED).  It will be obvious how
to generalize the results.

The Lagrangian for NCQED,
\begin{equation}
\label{eq:NCQED}
  \mathcal{L} = - \frac{1}{4} F_{\mu\nu}\ast F_{\mu\nu}
                + \bar\psi\ast(\ii\fmslash{D} - m)\ast\psi\,,
\end{equation}
with
\begin{subequations}
\begin{align}
  D_\mu &= \partial_\mu - \ii e A_\mu \\
  F_{\mu\nu} &= \frac{\ii}{e} [D_\mu\astcomma D_\nu]
              = \partial_\mu A_\nu - \partial_\nu A_\mu - \ii e [A_\mu\astcomma A_\nu]
\end{align}
\end{subequations}
is invariant under the gauge transformations
\begin{subequations}
\begin{align}
  \delta_\eta\psi &= \ii e \eta\ast\psi \\
  \delta_\eta\bar\psi &= -\ii e \bar\psi\ast\eta \\
  \delta_\eta A_\mu &= [D_\mu\astcomma\eta]\,,
\end{align}
\end{subequations}
where the $\ast$-commutator is defined by
\begin{equation}
  [f\astcomma g] = f\ast g - g\ast f\,.
\end{equation}
The gauge invariance can most easily be seen from the covariant
transformation rules
\begin{subequations}
\begin{align}
  \delta_\eta D_\mu &= - \ii e \delta_\eta A_\mu = \ii e[\eta\astcomma D_\mu]\\
  \delta_\eta(D_\mu\ast\psi) &= \ii e\eta\ast D_\mu\ast\psi\\
  \delta_\eta F_{\mu\nu} &= \ii e [\eta\astcomma F_{\mu\nu}]\,.
\end{align}
\end{subequations}
The cubic and quartic interactions in the Lagrangian~(\ref{eq:NCQED})
are given by
\begin{subequations}
\label{eq:NCQED-vertices}
\begin{align}
  \mathcal{L}_3 &=   e \bar\psi\ast\fmslash{A}\ast\psi
                   + e \ii\partial_\mu A_\nu \ast [A^\mu\astcomma A^\nu] \\
  \mathcal{L}_4 &= \frac{e^2}{4} [A_\mu\astcomma A_\nu] \ast [A^\mu\astcomma A^\nu]\,.
\end{align}
\end{subequations}
Taking into account the ambiguity~(\ref{eq:non-cyclic}) in assigning
the Moyal phases~$\varphi$ in TOPT, one has for the $e^+e^-\gamma$-vertex:
\begin{multline}
\label{eq:ordering}
   \bar\psi\ast\fmslash{A}\ast\psi \to
     (   c_1 \bar\psi_\alpha\ast A_\mu\ast\psi_\beta
       + c_2 A_\mu\ast\psi_\beta\ast\bar\psi_\alpha
       + c_3 \psi_\beta\ast\bar\psi_\alpha\ast A_\mu) \gamma^\mu_{\alpha\beta} \\
    \stackrel{\mathrm{F.T.}}{\longrightarrow}
      \sum_{i=1}^3 c_i \ee^{-\ii\varphi_i(\bar p,k,p)}
        \bar\psi(\bar p)\fmslash{A}(k)\psi(p)
\end{multline}
with arbitrary coefficients~$c_i$ obeying~$c_1+c_2+c_3=1$.  Here we
have introduced the notation $\varphi_l(k_1,k_2,k_3) = \varphi(k_l,k_m,k_n)$
for cyclical permutations~$\{l,m,n\}$ of~$\{1,2,3\}$.
With $p$ and $k$ incoming and $p'$ outgoing, the corresponding vertex
factor is given by
\begin{equation}
\label{eq:vertex-eeg}
  \parbox{40mm}{%
    \fmfframe(4,5)(4,5){%
      \begin{fmfgraph*}(30,20)
        \fmfleft{l,dl}
        \fmfright{r,dr}
        \fmftop{t}
        \fmf{boson,label=$p'-p=k$}{t,v}
        \fmf{fermion,label=$p$}{l,v}
        \fmf{fermion,label=$p'$,label.side=left}{v,r}
        \fmfdot{v}
      \end{fmfgraph*}}}
     = \ii e \gamma_\mu \sum_{i=1}^3 c_i \ee^{-\ii\varphi_i(-p',k,p)}
\end{equation}
While the phase factors in~(\ref{eq:ordering}) and~(\ref{eq:vertex-eeg}) are
fixed by TOPT, the choice of the coefficients~$c_i$ is ambiguous.  In the case of
the 3$\gamma$-vertex
\begin{equation}
\label{eq:vertex-ggg}
  \parbox{40mm}{%
    \fmfframe(4,5)(4,5){%
      \begin{fmfgraph*}(30,20)
        \fmfleft{l,dl}
        \fmfright{r,dr}
        \fmftop{t}
        \fmf{boson,label=$k_1$}{t,v}
        \fmf{boson,label=$k_2$}{l,v}
        \fmf{boson,label=$k_3$}{r,v}
        \fmfdot{v}
      \end{fmfgraph*}}}
    = \ii e V_{\mu_1,\mu_2,\mu_3}(k_1,k_2,k_3)
\end{equation}
the corresponding ambiguity in the Moyal phases leads to
\begin{multline}
\label{eq:dA[A,A]}
    c'_1 \ii\partial_\mu A_\nu\ast A^\mu\ast A^\nu
  + c'_2 A^\mu\ast A^\nu\ast \ii\partial_\mu A_\nu
  + c'_3 A^\nu\ast \ii\partial_\mu A_\nu\ast A^\mu \\
  - c'_1 \ii\partial_\mu A_\nu\ast A^\nu\ast A^\mu
  - c'_2 A^\nu\ast A^\mu\ast \ii\partial_\mu A_\nu
  - c'_3 A^\mu\ast \ii\partial_\mu A_\nu\ast A^\nu
\end{multline}
with $c'_1+c'_2+c'_3=1$. In general, the coefficients~$c'_i$ and~$c_i$
are different.

When Fourier transforming~(\ref{eq:dA[A,A]}),
one faces another ambiguity associated
with the derivative couplings. In covariant perturbation theory,
derivatives can be shifted by partial integration from one field to
the other fields at the same vertex, with energy-momentum conservation
insuring that the final result is independent of this choice.  In TOPT
however, one will get different results from a derivative acting on
one field and a derivative acting on the product of the other fields.
While in a local QFT the physical results can be shown to be
equivalent by using the equations of
motion~\cite{Grosse-Knetter:1993td}, this is not the case in a
non-commutative QFT with an infinite number of time derivatives.  In
section~\ref{sec:WI} we will use the WIs to derive an unambiguous
prescription for the choice of the momenta corresponding to the
derivatives after Fourier transformation.  These momenta are denoted
by~$\bar k_i$ for now.

With this convention the interaction~(\ref{eq:dA[A,A]}) yields the
vertex factor
\begin{multline}
\label{eq:V(A,A,A)}
  \ii V_{\mu_1,\mu_2,\mu_3}(k_1,k_2,k_3) = \\
    \sum_{i=1}^3 c'_i
      \left(\bar k_1^{\mu_2} g^{\mu_1\mu_3} - \bar k_1^{\mu_3} g^{\mu_1\mu_2}\right)
      \left(\ee^{-\ii\varphi_i(k_1,k_2,k_3)} - \ee^{-\ii\varphi_i(k_1,k_3,k_2)}\right)\\
    + \text{cyclic $\{1,2,3\}$}\,.
\end{multline}
where $k_{1,2,3}$ appearing in the Moyal phases are the on-shell
momenta of~TOPT. Defining
\begin{equation}
  \delta k = k_1 + k_2 + k_3 \,,
\end{equation}
the expression~(\ref{eq:V(A,A,A)}) can be written in the form
\begin{equation}
  \ii V_{\mu_1,\mu_2,\mu_3}(k_1,k_2,k_3) =
    \left(\bar k_1^{\mu_2} g^{\mu_1\mu_3} - \bar k_1^{\mu_3} g^{\mu_1\mu_2}\right)
        \mathcal{C}_{23}(k_1,k_2,k_3) + \text{cyclic $\{1,2,3\}$}
\end{equation}
with
\begin{subequations}
\begin{align}
  \mathcal{C}_{23}(k_1,k_2,k_3) &= 
     2 \sin(k_2\wedge k_3) C_1(k_1,k_2,k_3) + 2 C_0(k_1,k_2,k_3)\\
\intertext{and}
  C_1(k_1,k_2,k_3)
   &=   c'_1\ee^{-\ii k_1\wedge\delta k}
      + c'_2\ee^{ \ii k_1\wedge\delta k}
      + c'_3\cos(\delta k\wedge(k_2-k_3)) \\
  C_0(k_1,k_2,k_3)
   &= c'_3\cos(k_2\wedge k_3)\sin(\delta k\wedge(k_2-k_3))
\end{align}
\end{subequations}
Note that
\begin{equation}
  \mathcal{C}_{23}(k_1,k_2,k_3)\Bigr|_{\delta k=0}
    = 2 \sin(k_2\wedge k_3)
\end{equation}
as in NCQED for $\theta_{ij}\not=0$~\cite{Mariz:2002hd}.

\section{Ward Identities for Compton Scattering}
\label{sec:WI}
\label{sec:compton}
After adding the gauge-fixing term
\begin{equation}
\label{eq:gf}
  \mathcal{L}_{\text{g.f.}} =
    \delta_{\text{BRST}}
      \left(\bar c\ast\left(\frac{\xi}{2}B + \partial_\mu A^\mu\right)\right)
\end{equation}
with Faddeev-Popov ghosts~$c$, antighosts~$\bar c$ and a
Nakanishi-Lautrup field~$B$, to the Lagrangian~(\ref{eq:NCQED}), the
sum~$\mathcal{L}+\mathcal{L}_{\text{g.f.}}$ is invariant under the BRST
transformations~\cite{Soroush:2003pk}
\begin{subequations}
\label{eq:BRST}
\begin{align}
  \delta_{\text{BRST}}\psi &= \ii e c\ast\psi \\
  \delta_{\text{BRST}}\bar\psi &= -\ii e \bar\psi\ast c \\
\label{eq:delta(A)}
  \delta_{\text{BRST}}A_\mu &= [D_\mu\astcomma c] \\
  \delta_{\text{BRST}}c &= \ii [c\astcomma c] \\
  \delta_{\text{BRST}}\bar c &= B \\
  \delta_{\text{BRST}}B &= 0\,.
\end{align}
\end{subequations}
The invariance of the action under~(\ref{eq:BRST}) engenders relations
among Green functions of the theory, known as Slavnov-Taylor
identities~(STIs).  The prototype STI is derived from
\begin{equation}
   \Braket{0|\mathrm{T} \delta_{\text{BRST}} \Bigl(\bar c(x)
     \Phi_1(x_1)\Phi_2(x_2)\cdots\Phi_n(x_n) \Bigr)|0} = 0
\end{equation}
using the
equation of motion for~$B=-\partial_\mu A^\mu/\xi$:
\begin{multline}
\label{eq:STI}
  \frac{\partial}{\partial x^\mu}
    \Braket{0|\mathrm{T} A^\mu(x)
       \Phi_1(x_1)\Phi_2(x_2)\cdots\Phi_n(x_n)|0} = \\
  \xi \sum_{i} (\pm)
    \Braket{0|\mathrm{T} \bar c(x)
       \Phi_1(x_1)\cdots\delta_{\text{BRST}}\Phi_i(x_i)\cdots\Phi_n(x_n)|0}\,,
\end{multline}
where the sign of each summand is fixed by the anticommuting nature of
the BRST transformation.

In~(\ref{eq:BRST}) the BRST transforms of
the physical degrees of freedom~$\psi$, $\bar\psi$ and~$A_\mu$ (with
$\partial_\mu A^\mu=0$) are bilinear in these fields and the ghost~$c$.
Therefore, the contributions of these transforms to the STIs are cancelled
when matrix elements of physical fields are amputated on-shell.
As a consequence, the STIs~(\ref{eq:STI}) reduce on-shell to the WIs
\begin{equation}
  \frac{\partial}{\partial x^\mu}
    \Braket{0|\mathrm{T} A^\mu(x)
       \Phi_1(x_1)\Phi_2(x_2)\cdots\Phi_n(x_n)|0}_{\text{amputated, on-shell}} = 0
    \tag{\ref{eq:WI}}
\end{equation}
if all fields~$\Phi_j$ are either matter fields
or gauge fields with physical polarizations.

When any one of the WIs~(\ref{eq:WI}) is violated, the BRST
charge~$Q_{\text{BRST}}$ generating~(\ref{eq:BRST}) is not conserved
and it is impossible to construct a positive norm Hilbert space for
the physical asymptotic states from the cohomology of the BRST
operator using the
condition~$Q_{\text{BRST}}\Ket{\text{phys}}=0$.  Therefore, one cannot
not make physical sense of a gauge theory, unless the
WIs~(\ref{eq:WI}) are satisfied.

The simplest example for the violation of WIs in TOPT is provided by
Compton scattering~$e^-\gamma\to e^-\gamma$ or any of its crossed
variants.  The three contributing Feynman diagrams
\begin{subequations}
\begin{align}
  \parbox{30mm}{%
    \fmfframe(2,5)(2,5){%
      \begin{fmfgraph*}(26,15)
        \fmfleft{k1,p1}
        \fmfright{k2,p2}
        \fmflabel{$k_1$}{k1}
        \fmflabel{$k_2$}{k2}
        \fmflabel{$p_1$}{p1}
        \fmflabel{$p_2$}{p2}
        \fmf{fermion}{p1,v1}
        \fmf{fermion,tension=0.5,label=$q_s$,label.side=left}{v1,v2}
        \fmf{fermion}{v2,p2}
        \fmf{boson}{k1,v1}
        \fmf{boson}{v2,k2}
        \fmfdot{v1,v2}
      \end{fmfgraph*}}} &
       = \ii e^2\mathcal{M}^s_{s_1s_2,\mu_1\mu_2}(p_1,p_2,k_1,k_2) \\
  \parbox{30mm}{%
    \fmfframe(2,5)(2,5){%
      \begin{fmfgraph*}(26,15)
        \fmfleft{k1,p1}
        \fmfright{k2,p2}
        \fmflabel{$k_1$}{k1}
        \fmflabel{$k_2$}{k2}
        \fmflabel{$p_1$}{p1}
        \fmflabel{$p_2$}{p2}
        \fmf{fermion}{p1,v1}
        \fmf{fermion,tension=0.5,label=$q_u$,label.side=left}{v1,v2}
        \fmf{fermion}{v2,p2}
        \fmf{phantom}{k1,v1}
        \fmf{phantom}{v2,k2}
        \fmffreeze
        \fmf{boson}{k1,v2}
        \fmf{boson,rubout}{v1,k2}
        \fmfdot{v1,v2}
      \end{fmfgraph*}}} &
       = \ii e^2\mathcal{M}^u_{s_1s_2,\mu_1\mu_2}(p_1,p_2,k_1,k_2) \\
  \parbox{30mm}{%
    \fmfframe(2,5)(2,5){%
      \begin{fmfgraph*}(26,15)
        \fmfleft{k1,p1}
        \fmfright{k2,p2}
        \fmflabel{$k_1$}{k1}
        \fmflabel{$k_2$}{k2}
        \fmflabel{$p_1$}{p1}
        \fmflabel{$p_2$}{p2}
        \fmf{fermion}{p1,v1,p2}
        \fmf{boson,tension=0.5,label=$q_t$}{v1,v2}
        \fmf{boson}{k1,v2,k2}
        \fmfdot{v1,v2}
      \end{fmfgraph*}}} &
       = \ii e^2\mathcal{M}^t_{s_1s_2,\mu_1\mu_2}(p_1,p_2,k_1,k_2)\,.
\end{align}
\end{subequations}
yield
\begin{multline}
  k_1^{\mu_1} 
    \Braket{0|\mathrm{T} A_{\mu_1}(k_1)
       \left(\epsilon_{(\kappa)}^{\mu_2}(k_2)A_{\mu_2}(k_2)\right)
       \psi_1(p_1)\bar\psi_2(p_2)|0}_{\text{amputated, on-shell}}\\
    = \mathcal{W}_{(\kappa)}^s
    + \mathcal{W}_{(\kappa)}^u
    + \mathcal{W}_{(\kappa)}^t
\end{multline}
for photon polarizations~$\kappa=\pm$.
Starting with the $s$-channel and defining
\begin{subequations}
\begin{align}
  q_s &= p_1 + k_1 = p_2 + k_2 \\
  q_u &= p_1 - k_2 = p_2 - k_1 \\
  q_t &= p_2 - p_1 = k_1 - k_2\,,
\end{align}
\end{subequations}
we find 
\begin{subequations}
\label{eq:Ws}
\begin{equation}
  \mathcal{W}_{(\kappa)}^s
    = k_1^{\mu_1}\epsilon_{(\kappa)}^{\mu_2}(k_2)
        \mathcal{M}^s_{s_1s_2,\mu_1\mu_2}(p_1,p_2,k_1,k_2)
    = \mathcal{W}^{s,1}_{(\kappa)} + \mathcal{W}^{s,0}_{(\kappa)}
\end{equation}
with
\begin{align}
  \mathcal{W}^{s,1}_{(\kappa)} 
    &= \sum_{i,j=1}^3 c_ic_j R^{s,ij}(q_s^{(+)},q_s^{(-)})
       \bar u_{s_2}(p_2) \fmslash{\epsilon}_{(\kappa)} (k_2) u_{s_1}(p_1)\\
  \mathcal{W}^{s,0}_{(\kappa)} &=
     - \sum_{i,j=1}^3 c_ic_j R_-^{s,ij}(q_s^{(+)},q_s^{(-)})
       \frac{1}{q_{s,0}^{(+)}} 
       \bar u_{s_2}(p_2) \fmslash{\epsilon}_{(\kappa)} (k_2) \gamma_0 \fmslash{k}_1 u_{s_1}(p_1)\,.
\end{align}
and, using~(\ref{eq:R}) and~(\ref{eq:R+/-}),
\begin{align}
   R^{s,ij}(q_s^{(+)},q_s^{(-)}) &=
          \sum_{\lambda=\pm}
             \ee^{-\ii\varphi_i(-p_2,-k_2,q_s^{(\lambda)})}
             \ee^{-\ii\varphi_j(-q_s^{(\lambda)},k_1,p_1)}
             \frac{1}{2}\left(1+\frac{q_{s,0}}{q^{(\lambda)}_{s,0}}\right) \\
   R_-^{s,ij}(q_s^{(+)},q_s^{(-)}) &=
     \sum_{\lambda=\pm}
        \ee^{-\ii\varphi_i(-p_2,-k_2,q_s^{(\lambda)})}
        \ee^{-\ii\varphi_j(-q_s^{(\lambda)},k_1,p_1)}  \frac{\lambda}{2} \,.
\end{align}
\end{subequations}
Here, the term~$\mathcal{W}^{s,0}_{(\kappa)}$
remains after using the equation of motion for cancelling the electron
propagator
\begin{equation}
  \frac{1}{\fmslash{q}_s-m+\ii\varepsilon} \fmslash{k}_1 u_{s_1}(p_1)
    = \frac{1}{\fmslash{q}_s-m+\ii\varepsilon}
         (\fmslash{p}_1 + \fmslash{k}_1 - m) u_{s_1}(p_1)
    = u_{s_1}(p_1)\,.
\end{equation}
Similarly for the $u$-channel, one gets
\begin{subequations}
\label{eq:Wu}
\begin{equation}
  \mathcal{W}^u_{(\kappa)}
    = k_1^{\mu_1}\epsilon_{(\kappa)}^{\mu_2}(k_2)
        \mathcal{M}^u_{s_1s_2,\mu_1\mu_2}(p_1,p_2,k_1,k_2)
    = \mathcal{W}^{u,1}_{(\kappa)} + \mathcal{W}^{u,0}_{(\kappa)}
\end{equation}
with
\begin{align}
  \mathcal{W}^{u,1}_{(\kappa)}
    &= - \sum_{i,j=1}^3 c_ic_j R^{u,ij}(q_u^{(+)},q_u^{(-)})
          \bar u_{s_2}(p_2) \fmslash{\epsilon}_{(\kappa)} (k_2) u_{s_1}(p_1) \\
  \mathcal{W}^{u,0}_{(\kappa)}
    &= - \sum_{i,j=1}^3 c_ic_j R_-^{u,ij}(q_u^{(+)},q_u^{(-)})
         \frac{1}{q_{u,0}^{(+)}} 
         \bar u_{s_2}(p_2) \fmslash{k}_1 \gamma_0 \fmslash{\epsilon}_{(\kappa)} (k_2) u_{s_1}(p_1)
\end{align}
and
\begin{align}
   R^{u,ij}(q_u^{(+)},q_u^{(-)}) &=
            \sum_{\lambda=\pm}
               \ee^{-\ii\varphi_i(-p_2,k_1,q_u^{(\lambda)})}
               \ee^{-\ii\varphi_j(-q_u^{(\lambda)},-k_2,p_1)}
          \frac{1}{2}\left(1+\frac{q_{u,0}}{q^{(\lambda)}_{u,0}}\right) \\
   R_-^{u,ij}(q_u^{(+)},q_u^{(-)}) &=
     \sum_{\lambda=\pm}
        \ee^{-\ii\varphi_i(-p_2,k_1,q_u^{(\lambda)})}
        \ee^{-\ii\varphi_j(-q_u^{(\lambda)},-k_2,p_1)}
        \frac{\lambda}{2}\,.
\end{align}
\end{subequations}
Finally, the $t$-channel contribution is given by
\begin{multline}
\label{eq:Wt}
  \mathcal{W}^t_{(\kappa)} =
   k_1^{\mu_1}\epsilon_{(\kappa)}^{\mu_2}(k_2)
     \mathcal{M}^t_{s_1s_2,\mu_1\mu_2}(p_1,p_2,k_1,k_2) \\
    =  \sum_{i=1}^3 c_i
          \sum_{\lambda=\pm}
          \ee^{-\ii\varphi_i(-p_2,q_t^{(\lambda)},p_1)}
          \frac{1}{2}\left(1+\frac{q_{u,0}}{q^{(\lambda)}_{u,0}}\right) \times\\
          \bar u_{s_2}(p_2) \gamma^{\mu_3} u_{s_1}(p_1)
          \frac{1}{q_t^2}
          V_{\mu_1\mu_2\mu_3}(k_1,-k_2,-q_t^{(\lambda)})
          k_1^{\mu_1}\epsilon_{(\kappa)}^{\mu_2}(k_2)
\end{multline}
where the additional phases are absorbed in the 3$\gamma$ vertex
factor~$V$.

In the case of vanishing time-like noncommutativity,
i.\,e.~$\theta_{0i}=0$, we can use~(\ref{eq:Moyal-reduced}) to remove
all dependence of the Moyal phases on the internal momenta.  Then the
sums over~$i$, $j$ and~$\lambda$ become trivial, yielding
\begin{subequations}
\begin{align}
  \mathcal{W}^s_{(\kappa)}
    &= \ee^{-\ii\varphi(-p_2,-k_2,k_1,p_1)}
          \bar u_{s_2}(p_2) \fmslash{\epsilon}_{(\kappa)} (k_2) u_{s_1}(p_1) \\
  \mathcal{W}^u_{(\kappa)}
    &= - \ee^{-\ii\varphi(-p_2,k_1,-k_2,p_1)}
          \bar u_{s_2}(p_2) \fmslash{\epsilon}_{(\kappa)} (k_2) u_{s_1}(p_1) \\
  \mathcal{W}^t_{(\kappa)}
    &= \ee^{-\ii\varphi(p_1,-p_2)}
          \bar u_{s_2}(p_2) \gamma^{\mu_3} u_{s_1}(p_1)
          \frac{1}{q_t^2}
          V_{\mu_1\mu_2\mu_3}(k_1,-k_2,-q_t)
          k_1^{\mu_1}\epsilon_{(\kappa)}^{\mu_2}(k_2)\,.
\end{align}
\end{subequations}
Furthermore,
the $s$- and $u$-channel contributions can be combined since the 
overall energy conservation makes the phases cyclically symmetric
\begin{equation}
\label{eq:WI(space-like)}
  \mathcal{W}^s_{(\kappa)} + \mathcal{W}^u_{(\kappa)}
     = -2\ii\sin\left(k_1\wedge k_2\right) \ee^{\ii p_1\wedge p_2}
          \bar u_{s_2}(p_2) \fmslash{\epsilon}_{(\kappa)} (k_2) u_{s_1}(p_1)\,.
\end{equation}
Using
\begin{equation}
    k_1^{\mu_1}\epsilon_{(\kappa)}^{\mu_2}(k_2)
      V_{\mu_1\mu_2\mu_3}(k_1,-k_2,-q_t)
        = \ii \left( q_t^2 \fmslash{\epsilon}_{(\kappa)} (k_2)
                 - (q_t\epsilon_{(\kappa)} (k_2)) q_{t,\mu_3}\right)
                   2 \sin\left(k_1\wedge k_2\right) 
\end{equation}
one finally recovers $\mathcal{W}^s_{(\kappa)}+\mathcal{W}^t_{(\kappa)}
+\mathcal{W}^u_{(\kappa)}=0$~\cite{Mariz:2002hd}.

Returning to the general case,
the presence of the $1/q_t^2$~pole in~(\ref{eq:Wt}), together with its
absence in~(\ref{eq:Ws}) and~(\ref{eq:Wu}), shows that,
for any cancellation between $\mathcal{W}^s_{(\kappa)}+\mathcal{W}^u_{(\kappa)}$
and~$\mathcal{W}^t_{(\kappa)}$ to take place, we must have
\begin{equation}
\label{eq:WI(V)}
  k_1^{\mu_1}\epsilon_{(\kappa)}^{\mu_2}(k_2)
    V_{\mu_1\mu_2\mu_3}(k_1,-k_2,-q_t^{(\lambda)})
      = \alpha_1 q_t^2 \epsilon_{(\kappa),\mu_3}
            + \alpha_2 q_{t,\mu_3}\,,
\end{equation}
where the $\alpha_2$-term is allowed because it will not contribute due to
current conservation, i.\,e.~$\bar u(p_2) \fmslash{q}_t u(p_1)=0$.  In order
to determine the consequences of~(\ref{eq:WI(V)}), we can make the
general \textit{ansatz}
\begin{multline}
  \tilde V_{\mu_1\mu_2\mu_3}(b_1,b_2,b_3|k_1,k_2,k_3) =
      (b_1\bar k_{1,\mu_3} - b_2\bar k_{2,\mu_3})g_{\mu_1\mu_2} \\
    + (b_2\bar k_{2,\mu_1} - b_3\bar k_{3,\mu_1})g_{\mu_2\mu_3}
    + (b_3\bar k_{3,\mu_2} - b_1\bar k_{1,\mu_2})g_{\mu_3\mu_1}
\end{multline}
where the coefficients~$b_i$ can contain momentum dependent phase
factors.  For the process at hand, we have~$\bar k_1=k_1$ and~$\bar
k_2=k_2$ since they are
external on-shell momenta ($k_1^2=k_2^2=0$).
Using~$\epsilon_{(\kappa)}^\mu(k_2) k_{2,\mu}=0$, we obtain
\begin{multline}
\label{eq:b/deltak}
   k_1^{\mu_1}\epsilon_{(\kappa)}^{\mu_2}(k_2)
     \tilde V_{\mu_1\mu_2\mu_3}(b_1,b_2,b_3|k_1,k_2,k_3) = \\
   \frac{b_3 + b_2}{2} \bar k_3^2 \epsilon_{(\kappa),\mu_3}(k_2)
            - b (\bar k_3\epsilon_{(\kappa)}(k_2)) \bar k_{3,\mu_3}
      + (\bar k_3\epsilon_{(\kappa)}(k_2))
           (b_3 k_{1} + b_2 k_{2} + b\bar k_{3})_{\mu_3} \\
      - b_2 (\delta\bar k\epsilon_{(\kappa)}) k_{2,\mu_3}
      + (\delta\bar k(b_3 k_2-b_2\bar k_3)) \epsilon_{(\kappa),\mu_3}
      + \frac{b_2 - b_3}{2} (\delta\bar k)^2 \epsilon_{(\kappa),\mu_3}\,,
\end{multline}
where the term proportional to~$b$ has been added and subtracted.  The
first term in the right hand side of~(\ref{eq:b/deltak}) corresponds
to the term proportional to~$\alpha_1$ in~(\ref{eq:WI(V)}) and cancels
the pole.  Like the term proportional to~$\alpha_2$
in~(\ref{eq:WI(V)}), the second term will not contribute to the WI.
The remaining terms have to vanish.  This gives us two conditions for
the momenta in the derivative couplings
\begin{subequations}
\label{eq:derivative-couplings}
\begin{align}
  \delta\bar k = k_1 + k_2 + \bar k_3 &= 0 \\
  b_3 k_1 + b_2 k_2 + b \bar k_3 &=0\,,
\end{align}
\end{subequations}
which can be satisfied simultaneously for $k_i\not=0$, if and only if
\begin{equation}
  b_3 = b_2 = b\,.
\end{equation}
Therefore these momenta must satisfy energy-momentum conservation and
cannot be the momenta in~TOPT that do not conserve energy at the
vertices.

At this point, we could proceed by attempting to solve the
WI $\mathcal{W}^s_{(\kappa)}+\mathcal{W}^u_{(\kappa)}
+\mathcal{W}^t_{(\kappa)}=0$ explicitly, using the
conditions collected so far.  However, it will turn out in the next
section that the general structure
of the phase factors in~(\ref{eq:Ws}), (\ref{eq:Wu}) and~(\ref{eq:Wt})
provides enough constraints for deciding whether~(\ref{eq:WI}) can be
solved.

\section{Mismatched Phases}
\label{sec:no-go}
It turns out that already processes involving two gauge bosons and two
matter fields as discussed in section~\ref{sec:compton} suffice to
demonstrate the violation of~WIs in~TOPT.

Since all factors multiplying~$c_i c_j$ and~$c_i c'_j$ in the
contributions from the three channels~(\ref{eq:Ws}), (\ref{eq:Wu})
and~(\ref{eq:Wt}) are non-zero, the coefficients~$c_i$ and~$c'_i$ for
the $e^+e^-\gamma$- and 3$\gamma$-vertices must be chosen such that
the corresponding combination of phase factors vanishes to get
$\mathcal{W}^s_{(\kappa)}+\mathcal{W}^u_{(\kappa)}+\mathcal{W}^t_{(\kappa)}=0$.

The overall momentum conservation $p_1+k_1=p_2+k_2$ constrains the
Moyal phases.  Taking these constraints into account, parameterizing
the violation of energy conservation at the vertices by
\begin{equation}
  \delta q_s^{(\lambda)} = q_s^{(\lambda)} - p_1 - k_1 = q_s^{(\lambda)} - p_2 - k_2
\end{equation}
and using an obvious notation with respect to the indices~$i$ and~$j$,
we find in the $s$-channel
\begin{multline}
\label{eq:phases-s}
  \ee^{-\ii(\varphi_i(-p_2,-k_2,q_s^{(\lambda)})+\varphi_j(-q_s^{(\lambda)},k_1,p_1))}\\
    = \ee^{-\ii p_1\wedge p_2} \ee^{\ii k_1\wedge k_2}
        \begin{bmatrix}
            1
          & \ee^{-2\ii\delta q_s^{(\lambda)} \wedge p_2}
          & \ee^{-2\ii\delta q_s^{(\lambda)} \wedge q_s}\\
            \ee^{ 2\ii\delta q_s^{(\lambda)} \wedge q_s}
          & \ee^{ 2\ii\delta q_s^{(\lambda)} \wedge k_2}
          & 1 \\
            \ee^{ 2\ii\delta q_s^{(\lambda)} \wedge p_1}
          & \ee^{-2\ii\delta q_s^{(\lambda)} \wedge q_t}
          & \ee^{-2\ii\delta q_s^{(\lambda)} \wedge k_1}
        \end{bmatrix}\,.
\end{multline}
Analogously for the $u$-channel ($k_1\leftrightarrow-k_2$), we obtain 
\begin{multline}
\label{eq:phases-u}
  \ee^{-\ii(\varphi_i(-p_2,k_1,q_u^{(\lambda)})+\varphi_j(-q_u^{(\lambda)},-k_2,p_1))}\\
    = \ee^{-\ii p_1\wedge p_2} \ee^{-\ii k_1\wedge k_2}
        \begin{bmatrix}
            1
          & \ee^{-2\ii\delta q_u^{(\lambda)} \wedge p_2}
          & \ee^{-2\ii\delta q_u^{(\lambda)} \wedge q_u} \\
            \ee^{ 2\ii\delta q_u^{(\lambda)} \wedge q_u}
          & \ee^{-2\ii\delta q_u^{(\lambda)} \wedge k_1}
          & 1 \\
            \ee^{ 2\ii\delta q_u^{(\lambda)} \wedge p_1}
          & \ee^{-2\ii\delta q_u^{(\lambda)} \wedge q_t}
          & \ee^{ 2\ii\delta q_u^{(\lambda)} \wedge k_2}
        \end{bmatrix}
\end{multline}
with
\begin{equation}
   \delta q_u^{(\lambda)} = q_u^{(\lambda)} - p_1 + k_2 = q_u^{(\lambda)} - p_2 + k_1\,.
\end{equation}
Finally in the $t$-channel, with
\begin{equation}
   \delta q_t^{(\lambda)} = q_t^{(\lambda)} - p_1 + p_2 = q_t^{(\lambda)} + k_1 - k_2\,,
\end{equation}
the phases factors are a combination of
\begin{multline}
\label{eq:phases-t}
  \ee^{-\ii(\varphi_i(-p_2,q_t^{(\lambda)},p_1)+\varphi_j(k_1,-k_2,-q_t^{(\lambda)}))}\\
    = \ee^{-\ii p_1\wedge p_2}\ee^{\ii k_1\wedge k_2}
        \begin{bmatrix}
            \ee^{-2\ii\delta q_t^{(\lambda)} \wedge p_1}
          & \ee^{-2\ii\delta q_t^{(\lambda)} \wedge q_s}
          & \ee^{-2\ii\delta q_t^{(\lambda)} \wedge p_2} \\
            1
          & \ee^{-2\ii\delta q_t^{(\lambda)} \wedge k_1}
          & \ee^{-2\ii\delta q_t^{(\lambda)} \wedge q_t} \\
            \ee^{ 2\ii\delta q_t^{(\lambda)} \wedge q_t}
          & \ee^{-2\ii\delta q_t^{(\lambda)} \wedge k_2}
          & 1
        \end{bmatrix}
\end{multline}
and the same terms with~$k_1\leftrightarrow k_2$.  The phase
factor~$\ee^{-\ii p_1\wedge p_2}$ is common to all contributions.  The
factors~$\ee^{\pm\ii k_1\wedge k_2}$ must combine to a
factor~$\sin(k_1\wedge k_2)$ as in~(\ref{eq:WI(space-like)}) in order
to preserve the~WI.  This is not possible because of the remaining
phase factors that depend on the TOPT momenta~$\delta
q_s^{(\lambda)}$, $\delta q_u^{(\lambda)}$ and~$\delta q_t^{(\lambda)}$.

A simple example can serve as illustration: the dependence of the
phases on the external momenta in the $s$-channel (\ref{eq:phases-s})
and $u$-channel (\ref{eq:phases-u}) is only the same for the factors
multiplying~$c_1 c_1$, where the phases from TOPT have cancelled
altogether by the overall energy conservation. Choosing $c_1=1$ and
$c_2=c_3=0$ would turn off all other phases in~(\ref{eq:phases-s})
and~(\ref{eq:phases-u}), but then the factors multiplying~$c_1 c'_j$
in the $t$-channel~(\ref{eq:phases-t}) contain phases that depend
on~$\delta q_t^{(\lambda)}$ in a way that cannot be cancelled by other
factors in the matrix elements.

In phenomenological applications, the violation of the WI manifests
itself in negative results for cross sections calculated with
covariant polarization sums
\begin{equation}
   \sum_{\sigma=\pm} \epsilon_{\sigma}^\mu{\epsilon_{\sigma}^\nu}^*
     \to - g^{\mu\nu}
\end{equation}
and including the ghost diagram
\begin{equation}
  \parbox{30mm}{%
    \fmfframe(2,5)(2,5){%
      \begin{fmfgraph*}(26,15)
        \fmfleft{k1,p1}
        \fmfright{k2,p2}
        \fmflabel{$k_1$}{k1}
        \fmflabel{$k_2$}{k2}
        \fmflabel{$p_1$}{p1}
        \fmflabel{$p_2$}{p2}
        \fmf{fermion}{p1,v1,p2}
        \fmf{boson,tension=0.5,label=$q_t$}{v1,v2}
        \fmf{ghost}{k1,v2,k2}
        \fmfdot{v1,v2}
      \end{fmfgraph*}}}
\end{equation}
with the $\bar cc\gamma$~vertex given by the Faddeev-Popov
Lagrangian~$\mathcal{L}_{\text{g.f.}}$~(\ref{eq:gf}).
The pragmatic way of circumventing this problem by summing only over
physical polarizations could only postpone the problem, because the
negative tree level cross section will reappear as violations of the
cutting rules at one-loop level.

Calculations in an effective field theory approach to
NCGT~\cite{Wess:pr}, that consider only terms of a given finite
order in~$\theta_{i0}$ are not affected by the problems discussed in
this article.

\section{Conclusions}
We have shown that it is impossible to construct interaction vertices
for time-like Noncommutative Gauge Theory~(NCGT) in Time-Ordered
Perturbation Theory~(TOPT), in the form proposed
in~\cite{Liao:2002xc}, that lead to scattering amplitudes with
external gauge bosons which satisfy basic Ward
Identities~(WIs)~(\ref{eq:WI}).  Our arguments do not depend on
detailed features of the theory, but follow from 
phase factors that spoil the WIs.  We have worked in NCQED to simplify
some notations, but it is obvious that our arguments carry over to any
$\mathrm{U}(N)$ time-like NCGT.

Since the definition of a physical positive norm Hilbert space for
gauge theories rests on STIs and WIs, we have to conclude that TOPT
cannot be used to cure the unitarity problem of time-like NCGTs.  It
appears that only a prescription that
abrogates the relation between amputated Green functions and
scattering amplitudes, as expressed by the LSZ reduction formulae,
could rescue 
TOPT for time-like NCGT.  At present, we have no suggestion for such a
prescription.  Since it is reasonable to insist that observed
asymptotic states are 
described by a commuting field theory, we have doubts that such a
radical modification exists.  It is interesting to note in this
context, that the standard reduction formulae have also been questioned for
time-like NCQFT from another perspective~\cite{Chaichian:2003hx}.

An investigation of the approach of~\cite{Wess:pr} in the case of
time-like NCGT will be the subject of future research.

\subsection*{Acknowledgments}
We thank Christian Schwinn and Klaus Sibold for useful discussions.
This research is supported by Bundesministerium f\"ur Bildung und
Forschung Germany, grants 05HT1RDA/6 and 05HT1WWA/2.


\end{fmffile}

\begin{thebibliography}{99}
\bibitem{Douglas:2001ba}
M.~R.~Douglas and N.~A.~Nekrasov,
Rev.\ Mod.\ Phys.\  {\bf 73} (2001) 977
[arXiv:hep-th/0106048].
\bibitem{Alvarez-Gaume:2003mb}
L.~Alvarez-Gaume and M.~A.~Vazquez-Mozo,
Nucl.\ Phys.\ B {\bf 668} (2003) 293
[arXiv:hep-th/0305093].
\bibitem{Gomis:2000zz}
J.~Gomis and T.~Mehen,
Nucl.\ Phys.\ B {\bf 591} (2000) 265
[arXiv:hep-th/0005129].
\bibitem{Liao:2002xc}
Y.~Liao and K.~Sibold,
Eur.\ Phys.\ J.\ C {\bf 25} (2002) 469
[arXiv:hep-th/0205269].
\bibitem{Liao:2002pj}
Y.~Liao and K.~Sibold,
Eur.\ Phys.\ J.\ C {\bf 25} (2002) 479
[arXiv:hep-th/0206011].
\bibitem{Liao:2002kd}
Y.~Liao and C.~Dehne,
Eur.\ Phys.\ J.\ C {\bf 29} (2003) 125
[arXiv:hep-ph/0211425].
\bibitem{Wess:pr}
J.~Wess,
Commun.\ Math.\ Phys.\  {\bf 219} (2001) 247.
\bibitem{Grosse-Knetter:1993td}
C.~Grosse-Knetter,
Phys.\ Rev.\ D {\bf 49} (1994) 6709
[arXiv:hep-ph/9306321].
\bibitem{Mariz:2002hd}
T.~Mariz, C.~A.~de S. Pires and R.~F.~Ribeiro,
[arXiv:hep-ph/0211416].
\bibitem{Soroush:2003pk}
M.~Soroush,
Phys.\ Rev.\ D {\bf 67} (2003) 105005
[arXiv:hep-th/0302179].
\bibitem{Chaichian:2003hx}
M.~Chaichian, M.~N.~Mnatsakanova, A.~Tureanu and Y.~S.~Vernov,
HIP-2003-37-TH,
[arXiv:hep-th/0306158].
\end{thebibliography}
\end{document}